\documentclass[final,5p,times,twocolumn]{elsarticle}

\usepackage{amssymb}

\usepackage{amsmath}
\usepackage{color}
\usepackage{csquotes}
\usepackage{footmisc}

\journal{Phys. Lett. B}

\begin{document}
\begin{frontmatter}
		
\author[Colmar]{A.~Albert}
\author[UPC]{M.~Andr\'e}
\author[Genova]{M.~Anghinolfi}
\author[Erlangen]{G.~Anton}
\author[UPV]{M.~Ardid}
\author[CPPM]{J.-J.~Aubert}
\author[APC]{T.~Avgitas}
\author[APC]{B.~Baret}
\author[IFIC]{J.~Barrios-Mart\'{\i}}
\author[LAM]{S.~Basa}
% ...

\address[Colmar]{\scriptsize{GRPHE - Universit\'e de Haute Alsace - Institut universitaire de technologie de Colmar, 34 rue du Grillenbreit BP 50568 - 68008 Colmar, France}}
\address[UPC]{\scriptsize{Technical University of Catalonia, Laboratory of Applied Bioacoustics, Rambla Exposici\'o,08800 Vilanova i la Geltr\'u,Barcelona, Spain}}
\address[Genova]{\scriptsize{INFN - Sezione di Genova, Via Dodecaneso 33, 16146 Genova, Italy}}
\address[Erlangen]{\scriptsize{Friedrich-Alexander-Universit\"at Erlangen-N\"urnberg, Erlangen Centre for Astroparticle Physics, Erwin-Rommel-Str. 1, 91058 Erlangen, Germany}}
\address[UPV]{\scriptsize{Institut d'Investigaci\'o per a la Gesti\'o Integrada de les Zones Costaneres (IGIC) - Universitat Polit\`ecnica de Val\`encia. C/  Paranimf 1 , 46730 Gandia, Spain.}}
\address[CPPM]{\scriptsize{Aix-Marseille Universit\'e, CNRS/IN2P3, CPPM UMR 7346, 13288 Marseille, France}}
\address[APC]{\scriptsize{APC, Universit\'e Paris Diderot, CNRS/IN2P3, CEA/IRFU, Observatoire de Paris, Sorbonne Paris Cit\'e, 75205 Paris, France}}
\address[IFIC]{\scriptsize{IFIC - Instituto de F\'isica Corpuscular (CSIC - Universitat de Val\`encia) c/ Catedr\'atico Jos\'e Beltr\'an, 2 E-46980 Paterna, Valencia, Spain}}
\address[LAM]{\scriptsize{LAM - Laboratoire d'Astrophysique de Marseille, P\^ole de l'\'Etoile Site de Ch\^ateau-Gombert, rue Fr\'ed\'eric Joliot-Curie 38,  13388 Marseille Cedex 13, France}}
% ...
\address[Genova-UNI]{\scriptsize{Dipartimento di Fisica dell'Universit\`a, Via Dodecaneso 33, 16146 Genova, Italy}}
% ...
		

\title{Results from the search for dark matter in the Milky Way with 9 years of data of the
ANTARES neutrino telescope''}

\begin{abstract}
	
	Using data recorded with the ANTARES telescope from 2007 to 2015, a new search for dark matter annihilation in the Milky Way has been performed. Three halo models and five annihilation channels, 
	$\rm WIMP + WIMP \to b \bar b, W^+ W^-, \tau^+ \tau^-, \mu^{+} \mu^{-}$ and $\nu \bar{\nu}$, with WIMP masses ranging from 50~$\text{GeV}/\text{c}^2$ to 100~$\text{TeV}/\text{c}^2$, 
	were considered. No excess over the expected background was found, and limits on the thermally averaged annihilation cross section were set. 
	
\end{abstract}


\end{frontmatter}


\section{Introduction}
\label{intro}

A wide variety of observations supply evidence for the existence of dark matter (DM)~\cite{DM_rev,darkmatter}. Its nature, however, is so-far unknown, and attempts to elucidate it have given rise to a lively and varied research programme in physics. A common hypothesis is to consider dark matter to be made of new, unknown particles. The assumption that these particles are a thermal relic of the Big Bang leads to the conclusion that they are weakly interacting massive particles (WIMPs). 

Different approaches are used to search for these particles: production at particle accelerators~\cite{Mitsou}, direct detection of the recoil from collisions with nuclei~\cite{Cline} or indirect detection by means of the secondary particles that they produce when they decay or annihilate~\cite{Gaskins}. Most of the particles that have been put forward as WIMPs candidates annihilate in pairs and subsequently produce standard model particles, including neutrinos. Neutrino telescopes may play a paramount role in the search for WIMPs via their annihilation products, because of their particularly clean signals and low expected backgrounds. 

In this paper the results from the search for dark matter in the Milky Way  using data recorded with the ANTARES neutrino telescope from 2007 to 2015, with a total live time of 2102 days are presented. Only neutrinos detected via muons produced inside or around the detector are considered. Here and in the following \enquote{neutrino} means $\nu_\mu + \bar{\nu}_\mu$, unless stated otherwise.

In Section \ref{sec:2} it is presented how the neutrino flux can be derived from the annihilation of DM particles. The detector and the reconstruction method are described in Section \ref{sec:3}, while the new analysis methodology is explained in Section \ref{sec:4}. The results are presented in Section \ref{sec:5}.

Compared to work previously published~\cite{Ant_dmgc}, a considerably increased data sample is used and a maximum likelihood method or \enquote{unbinned method} is applied. In addition, more recent parameters for the DM halo in the Milky Way are used.

\section{Dark matter phenomenology}
\label{sec:2}

In this type of indirect search two important ingredients have to be considered: the amount and spatial distribution of dark matter in the source under consideration, and the energy spectra of the standard model particles produced by WIMP annihilation. These two features are to a large extent independent of each other. They are relevant for modelling the expected signal and enter into the analysis at different stages.

The signal spectra used for the analysis presented here were calculated using the code described in~\cite{Cirelli_spectra}. Spectra were obtained for five annihilation channels and 17 WIMP masses between 50~$\text{GeV}/\text{c}^2$ and 
100~$\text{TeV}/\text{c}^2$. These spectra take into account the effect of neutrino oscillations. In the following, the results for each annihilation channel are given assuming a 100\% branching ratio. The five annihilation channels are:

\begin{equation}
\rm WIMP + WIMP \to b \bar{b}, W^+ W^-, \tau^+ \tau^-, \mu^+ \mu^-, \nu_{\mu} \bar{\nu}_{\mu}. \label{channels}
\end{equation}

Of these channels, the $b \bar{b}$-channel produces the softest neutrino spectra, whilst the $\nu_\mu \bar{\nu}_\mu$-channel produces the hardest spectra. Although the $\nu_\mu \bar{\nu}_\mu$-channel is suppressed in many models, such as those with the WIMP being the lightest neutralino of supersymmetric models, it is included in this study in order to be as model independent as possible. 

The second ingredient, i.e.\ the amount and distribution of dark matter in the source, is described by the so-called J-Factor. The J-Factor, $J(\psi)$, is the integral of the dark matter density squared, $\rm \rho_{DM}^2$, over a line of sight at an angular separation $\psi$ from the centre of the source. The relative signal strength at an angular separation $\psi$ to the source is described by the expression $J(\psi) d\Omega(\psi)$. The J-Factor can be integrated over an observation window $\Delta \Omega$:

\begin{equation}
\rm J_{int}(\Delta \Omega)= \int_{\Delta \Omega} \int \rho^2_{DM} \cdot dl \cdot d\Omega. \label{J-def}
\end{equation}

$\mathrm{J_{int}}$ relates the thermally averaged annihilation cross--section $\langle\sigma \mathrm{v}\rangle$ to the neutrino flux $\Phi_{\nu_\mu + \bar{\nu}_{\mu}}$ via the following equation:

\begin{equation}
\rm \frac{d \Phi_{\nu_\mu + \bar{\nu}_{\mu}}}{dE_{\nu_{\mu} + \bar{\nu_{\mu}}}} = \frac{\langle\sigma \mathrm{v}\rangle}{8 \pi M_{WIMP}^{2}} \cdot \frac{dN_{\nu_\mu + \bar{\nu}_{\mu}}}{dE_{\nu_{\mu} + \bar{\nu}_{\mu}}} \cdot J_{int}(\Delta \Omega) , \label{flux-rel}
\end{equation}

\noindent where $\mathrm{ N_{\nu_\mu + \bar{\nu}_{\mu}}}$ is the average number of neutrinos in the energy bin $\mathrm{dE}_{\nu_{\mu} + \bar{\nu}_{\mu}}$ per WIMP annihilation, v is the WIMP velocity and $\mathrm{M_{WIMP}}$ is the WIMP mass. 

\begin{figure}[h!]
\centering
\includegraphics[width=0.45\textwidth]{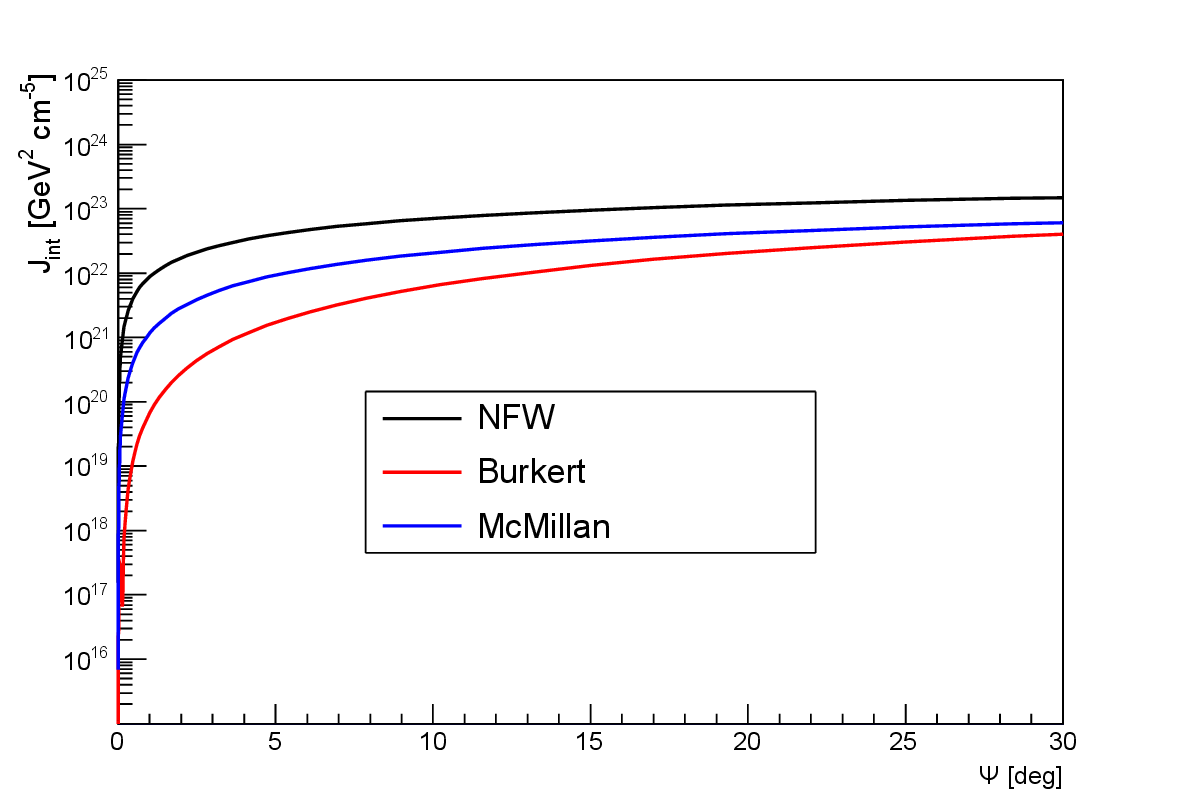}
 \caption{The integrated J-Factor, $J_{int}$, for a cone-shaped region $\Delta \Omega$ centred on the Galactic Centre with an opening angle $\Psi$. For the halo models the parameters from Table \ref{nesti_salucci_par} are used. The calculations are done using the code CLUMPY~\cite{CLUMPY}.}
\label{Jint}
\end{figure}

The shape of the J-Factor crucially depends on the halo model. In this analysis three models are used: the NFW~\cite{NFW}, the Burkert~\cite{burkert} model and the \enquote{McMillan}~\cite{mcmillan} profile. The parameters for these models are taken from~\cite{nesti_salucci} and~\cite{mcmillan} and are shown in Table \ref{nesti_salucci_par}. The McMillan profile is a variant of the Zhao profile~\cite{Zhao}, which treats one of the shape parameters, $\gamma$, as a free parameter and therefore is also referred to as the \enquote{$\gamma$ free} model. The optimum value of $\gamma$ for this model is $0.79 \pm 0.32$. The uncertainties on the halo profile parameters are not used in this analysis. In Figure \ref{Jint} the integrated J-Factors for the three models are shown. The NFW profile gives a larger total amount of dark matter that is also more concentrated in the core of the source than for the Burkert profile. This is due to the fact that the NFW profile is a so--called cuspy profile and diverges at the centre of the source, in contrast to the cored Burkert profile.

\begin{table}[h]
\begin{center}
\begin{tabular}{|l||c|c|c|c|}
\hline
Parameter & NFW & Burkert & McMillan \\
\hline 
\hline
$r_s$ $[kpc]$ & $16.1^{+17.0}_{-7.8}$ & $9.26^{+5.6}_{-4.2}$ &  $17.6 \pm 7.5$ \\
\hline
$\rho_{local}$ $[GeV/cm^3]$ & $0.471^{+0.048}_{-0.061}$ & $0.487^{+0.075}_{-0.088}$ & $0.390 \pm 0.034$  \\
\hline
\end{tabular}
\caption{Table of dark matter halo parameters for the Milky Way as taken from \cite{mcmillan} and \cite{nesti_salucci}. $\rho_{local}$ is the local density and $r_s$ is the scaling radius.}
\label{nesti_salucci_par}
\end{center}
\end{table}

\section{Simulation and reconstruction}
\label{sec:3}

The ANTARES neutrino telescope~\cite{antares} is installed at the bottom of the Mediterranean Sea, about 40 km from Toulon and about 2475 m below the sea surface. Being located in the Northern hemisphere ($42^\circ 48^\prime$ N, $6^\circ 10^\prime$ E) allows the ANTARES detector to directly observe the centre of the Milky Way, using the Earth as a shield against the background from atmospheric muons. 

ANTARES consists of 12, 450-m long, detector lines that are anchored to the seabed and kept vertical by buoys. Each line comprises 25 storeys with three 10--inch photomultipliers (PMTs)~\cite{PMT} per storey. The PMTs are housed inside pressure-resistant glass spheres~\cite{OM}.

The storeys also house the electronics to control the PMTs~\cite{DAQ} and a system to monitor the alignment of the lines~\cite{alignment}. For the synchronisation of the individual storeys a system of optical beacons~\cite{OB}, located at various points of the apparatus, is used~\cite{timing}.

In this analysis two muon track reconstruction strategies are used: $\Lambda$Fit and QFit. In the QFit strategy~\cite{BBFit} a  $\chi^2$-like quality parameter, Q, is minimised. Q is calculated from the squared difference between the expected and measured times of the detected photons, taking into account the effect of light absorption in the water~\cite{BBFit}. This strategy allows for the reconstruction of events with photon hits on only one line (single-line events).

$\Lambda$Fit~\cite{AAFit_official} maximises a likelihood ratio $\Lambda$ in a multistep process. The value of $\Lambda$ of the final iteration of this process is used as a measure of the quality of the reconstruction. In addition, the angular error estimate $\beta$ is used to define a cut employed to reduce the background.

The main background for analyses using muon tracks are atmospheric muons. Taking advantage of the absorption of the Earth that acts as an efficient shield against muons, most of this background can be rejected by accepting only upgoing-reconstructed muons in the analysis. Thanks to the detector's latitude, the centre of the Milky Way is efficiently observed, since it is below the horizon most of the time. To further reduce the background of atmospheric muons wrongly reconstructed as upgoing, cuts on the parameters that quantify the quality of the reconstruction (Q, $\Lambda$), and on the estimate of the angular error ($\beta$) are used, as specified in the next section. Atmospheric neutrinos are an additional but much smaller part of the background. However, unlike atmospheric muons, this background is irreducible, although the information of the energy and correlations with the source can help to discriminate it from the signal.

In order to evaluate the sensitivity of the search, Monte Carlo simulations, using a detailed detector response for each data run, have been performed \cite{km3}. Concerning the background, atmospheric neutrinos \cite{genhen} and muons \cite{MUPAGE} with energies ranging from 
10~$\text{GeV}/\text{c}^2$ to 100~$\text{TeV}/\text{c}^2$ have been simulated with the standard ANTARES simulation chain \cite{OM,biofouling,transmission}. From this simulation the detector resolution and acceptance is calculated for all five annihilation channels and for WIMP masses ranging from 
50~$\text{GeV}/\text{c}^2$ to 100~$\text{TeV}/\text{c}^2$.

In this paper, data taken from 2007 to 2015, corresponding to 2102 days of live time, was used. The agreement between the data and the simulation has been tested extensively for both reconstruction strategies. 

\section{Methodology}
\label{sec:4}

The maximum likelihood method is used to look for a signal of dark matter annihilation. The likelihood, which is a function of the number of signal events assumed to be present in the selected event sample, $\mathrm{n_s}$, is based on two probability distributions, S and B, which describe the behaviour of the signal and the background events, respectively, as a function of the relevant event variables. The likelihood is then maximised by varying $\mathrm{n_s}$. The statistical significance of the value obtained is extracted from the distribution of maximum likelihoods produced by generating pseudo-experiments, i.e.\ samples of events with known amounts of background and signal. The likelihood function used has the form

\begin{equation}
 \cal{L} \rm (n_s) =  e^{- (n_s+N_{\text{bg}})} \prod_{i = 1} ^{N_{tot}} \left( n_s S(\psi_i,N_{hit,i},\beta_i)
+N_{\text{bg}}B(\psi_i,N_{hit,i},\beta_i) \right),  \label{lik1}
\end{equation}

\noindent where $\mathrm{N_{bg}}$ is the expected number of background events, which is set equal to $\mathrm{ N_{tot}}$, the total number of reconstructed events. $\mathrm{n_s}$ is the variable that changes during the maximisation process. The two functions S and B depend  on: $\psi_i$, the angular distance of the $i$-th event to the centre of the Milky Way; $ N_{hit,i}$, the number of hits in the $i$-th event; and $ \beta_i$, the angular error estimate for the $i$-th event. The number of hits $\mathrm{N}_{\mathrm{hit,i}}$ is a proxy for the muon energy~\cite{ANT_ext}. 

In order to take the source extension into account, in S the non-integrated J-Factor, $\mathrm{J(\psi)}$, is used, smeared out with the point--spread function (PSF) assuming a 15\% systematic uncertainty on the angular resolution, which is the dominant systematic error from the detector in this analysis. This error is based on a 2.5 ns uncertainty in the timing of detected photon hits in ANTARES~\cite{PS}. By doing this, a combination of the PSF and the source morphology is obtained that is also used for generating signal events in the pseudo--experiments. 

Further uncertainties exist due to the choice of the halo model and the expected neutrino signal spectra. These uncertainties are studied by using different annihilation channels and halo profile functions in the analysis (see Figure \ref{sv_allchannel} and \ref{sv_model}).

A slightly modified likelihood function is defined for single--line events reconstructed with the QFit strategy:

\begin{equation}
 \cal{L} \rm (n_s) = e^{- (n_s+N_{\text{bg}})} \prod_{i = 1} ^{N_{tot}} \left( n_s \bar S(\theta_i,\bar{N}_{hit,i},Q_i)
+N_{\text{bg}}\bar B(\theta_i,\bar{N}_{hit,i},Q_i) \right) ,   \label{lik2}
\end{equation}

\noindent where $\mathrm{\bar{N}_{hit,i}}$ is the number of hits per storey (instead of the number of hits per PMT) used for the reconstruction, and $\theta_i$ is the difference in zenith angle between the $i$-th event and the centre of the Milky Way. $\mathrm{\bar S}$ and $\mathrm{\bar B}$ are the corresponding probability functions describing the signal and background distributions.

The likelihood functions are then studied using pseudo--experiments, which are generated from the distribution of background events from time--scrambled data and that of signal events from simulation. The signal events are generated by taking into account the angular resolution of the detector, the source morphology and the expected signal spectra. Ten thousand pseudo--experiments are simulated for each combination of WIMP mass, annihilation channel and reconstruction strategy, and for each considered value of signal events, $\mathrm{n_s}$. The maximum value considered for $\mathrm{n_s}$ is 80 for the QFit strategy and 120 (180) for the $\Lambda$Fit strategy using the NFW and McMillan (Burkert) profile. The maximum values were chosen because of differences in the amount of background in these cases. For each pseudo--experiment a test statistic (TS) is calculated: 

\begin{equation}
\rm TS =  log_{10}\left(\frac{{\cal L}(n_{opt})}{{\cal L}(0)}\right),
\end{equation}

\noindent where $\mathrm{n_{opt}}$ is the value of $\mathrm{n_s}$ that maximises the likelihood function. Since for a fixed signal strength the amount of detected events may vary, the TS distributions were combined using Poissonian weights producing new TS distributions. Sensitivities and limits are calculated following the approach suggested by Neyman~\cite{Neyman}. The 90\% C.L. sensitivity in terms of detected neutrino events, $\bar{\mu}_{90\%}$, is calculated as the average number of inserted signal events, which leads to TS values that are in 90\% of the cases above the median of the TS distribution for pure background. The 90\% C.L. limit in terms of detected neutrino events, $\mu_{90\%}$, is calculated by using the TS value of the unblinded data instead of the median of the background if this TS value is above the median; otherwise the limit is set to the sensitivity.

The event selection criteria, in particular the definition of the
cuts on Q and $\Lambda$ and the selection of the reconstruction strategy, have been optimised with the Model Rejection Factor method to obtain an unbiased cut selection for optimal sensitivities \cite{MRF}. The cut parameters have been tuned individually for each annihilation channel and several WIMP masses in the mass range under consideration, maintaining always a blind approach, i.e.\ with no access to the actual data.

It was found that for most combinations of WIMP mass and annihilation channels the optimum cuts are $Q < 0.7$ and  $\Lambda > -5.2$, respectively. Once $\bar{\mu}_{90\%}$ (the 90\% C.L. sensitivity on the average number
of signal events obtained from the likelihood function) is computed, the limits on the neutrino flux for a given mass $\mathrm{M_{WIMP}}$ and annihilation channel is calculated as

\begin{equation}
\overline{\Phi}_{\nu_\mu+\bar \nu_\mu,90\%} =  
\frac{\bar \mu_{90\%}(\mathrm{M_{WIMP},ch})}{\sum\limits_{\mathrm{i}} \overline{\mathcal{A}}^{\mathrm{i}}(\mathrm{M_{WIMP},ch}) \times \mathrm{T_{eff}^{i}}} \, ,
\end{equation}

\noindent where the index $\mathrm{i}$ denotes the periods with different detector configurations, $\mathrm{ch}$ the annihilation channel used and $\mathrm{ {T_{eff}^{i}}}$ the total corresponding livetime. In fact, throughout the considered 9 years, the number of available detector lines has changed from 5 to 12. The time span over which the number of available lines remains unchanged is defined as a particular detector configuration period. 
The effective area averaged over the neutrino energy, the {\textit{integrated acceptance}} $\mathrm{\overline{\mathcal{A}^i}_{eff}(\mathrm{M_{WIMP},ch})}$, is defined as:

\begin{gather}
\overline{\mathcal{A}^i}=\sum_{\nu,\bar{\nu}} \left( \frac{\int_{\mathrm{E_{\nu}^{th}}}^{\mathrm{M_{WIMP}}} \mathrm{A_{eff}^i}(E_{\nu,\bar{\nu}}) \, \left.\frac{dN_{\nu,\bar{\nu}}}{dE_{\nu,\bar{\nu}}}\right|_{\mathrm{ch,M_{WIMP}}} dE_{\nu,\bar{\nu}}}
{\int_{0}^{\mathrm{M_{WIMP}}}\left.\frac{dN_{\nu}}{dE_{\nu}}\right|_{\mathrm{ch,M_{WIMP}}} dE_{\nu} \,+\, \left.\frac{dN_{\bar{\nu}}}{dE_{\bar{\nu}}}\right|_{\mathrm{ch,M_{WIMP}}} dE_{\bar{\nu}}} \right)  \, , \label{Acc1}
\end{gather}

\noindent where $\mathrm{E_{\nu}^{th}}$ is the energy threshold for neutrino detection in ANTARES (approximatively 10 GeV), $\rm M_{WIMP}$ 
is the WIMP mass, $\rm dN_{\nu,\bar{\nu}}/dE_{\nu,\bar{\nu}}$ is the energy spectrum of the (anti-)neutrinos at the detector's location for annihilation channel $ch$ (see Equation 1) and WIMP mass $M_{WIMP}$, and $\rm A_{eff}(E_{\nu,\bar{\nu}})$ is the effective area of ANTARES as a function of the (anti-)neutrino energy. 

Due to their different cross-sections, the effective areas for neutrinos and anti-neutrinos are slightly different and therefore are considered separately. 
In addition, the fluxes of muon neutrinos and anti-neutrinos  are different and are convoluted with their respective efficiencies. 
The effective area for a detector configuration period is defined as the ratio between the neutrino event rate and the signal neutrino flux for a certain neutrino energy. It is calculated from simulation. 
In the first version of this paper, a too coarse numerical approximation\footnote{computed using the tabulated version of PPPC4 from  http://www.marcocirelli.net/PPPC4DMID.html\label{note1}} was used to fold the pre-binned 
tabulated values into our effective area. This lead to an overestimation of the integrated acceptance whose impact is proportional to the maximum neutrino energy considered, hence to the WIMP mass. 
Here, the same PPPC4 model was used but the convolutions over the neutrino energy have been produced analytically using the neutrino spectra obtained from the Mathematica functions\footref{note1}.  

\section{Results}
\label{sec:5}

The final results are obtained by comparing the TS value of the data, $\mathrm{TS_{obs}}$, to the TS distributions previously calculated under the blinded procedure. 

\begin{figure}[h!]
	\centering
	\includegraphics[width=0.49\textwidth]{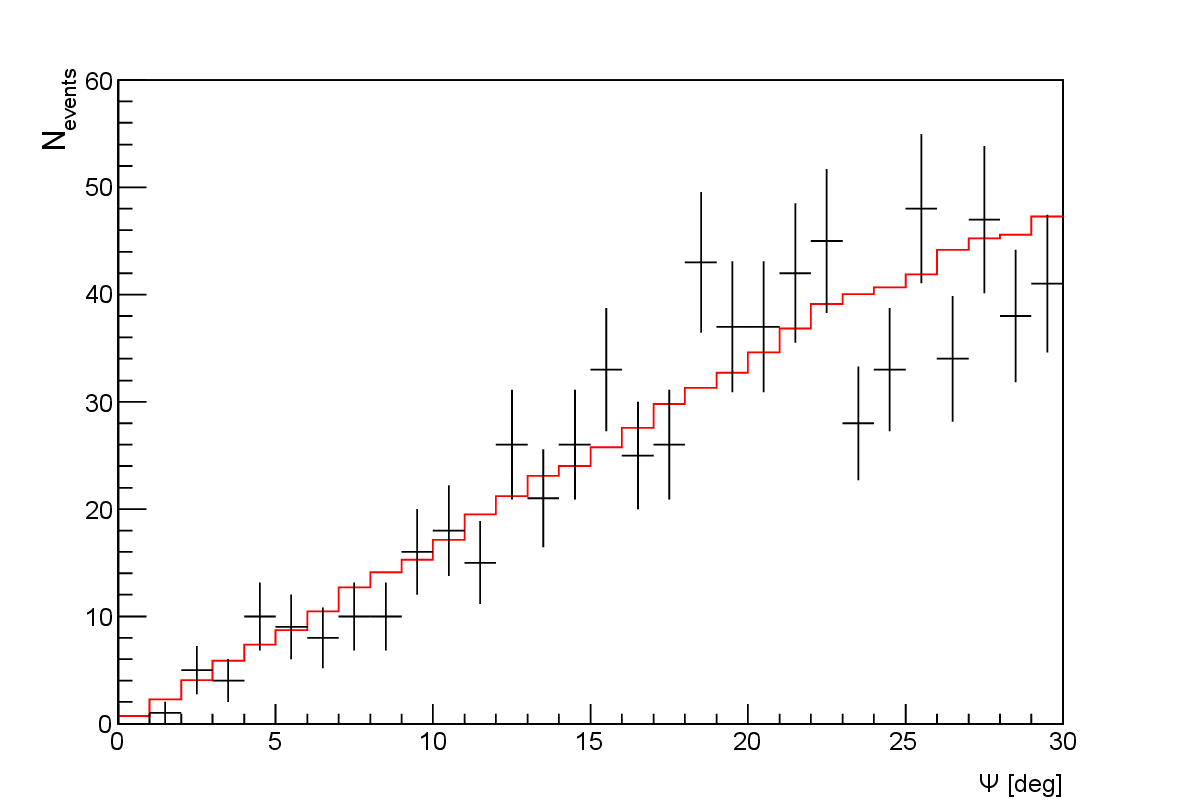}
	\caption{The number of events as a function of the distance to the Galactic Centre (crosses) in comparison to the background estimate (red line) for the $\Lambda$Fit reconstruction. For this plot a quality cut of $\Lambda > -5.2$ is used.}
	\label{Estimate}
\end{figure}

In Figure \ref{Estimate} a comparison between the unblinded data and the expected background is shown. No significant excess above the background can be seen, which is consistent with the fact that all the $\mathrm{TS_{obs}}$ values obtained are smaller than the medians of the corresponding background TS distributions. Since all background--like results should equally reject the considered dark matter model, upper limits have been set to the sensitivities calculated from the pseudo--experiments. 

\begin{figure}[h!]
\centering
\includegraphics[width=0.49\textwidth]{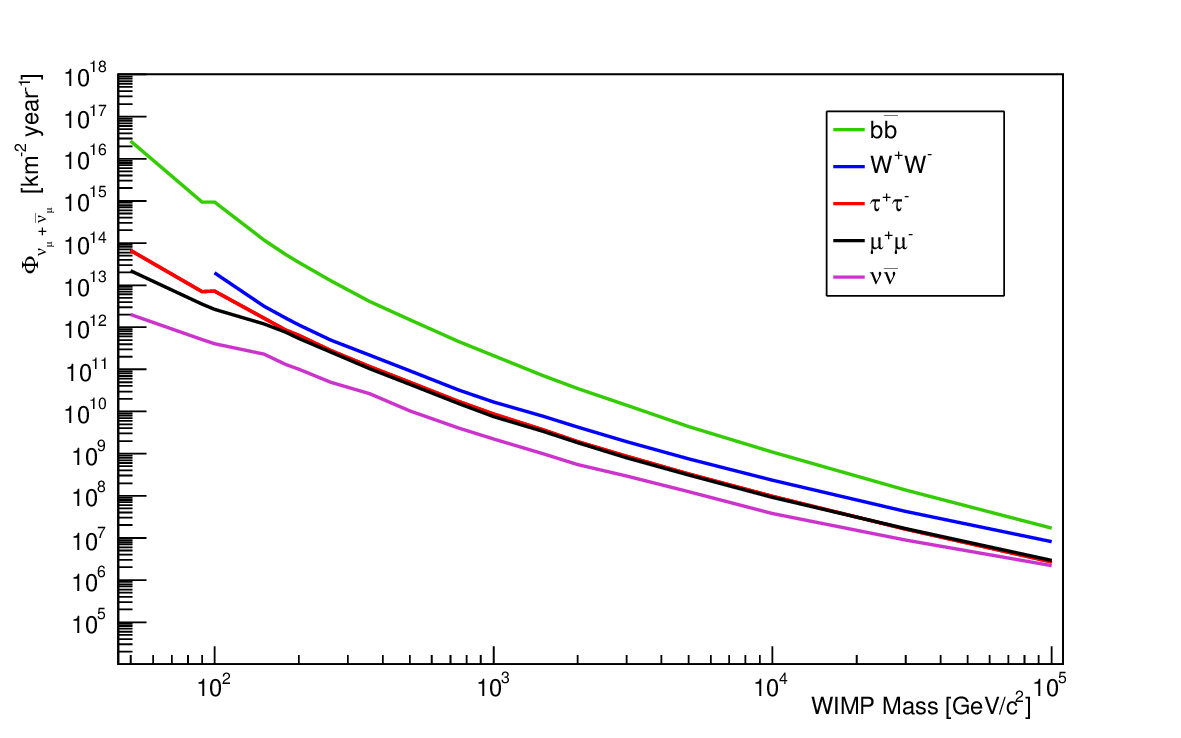} 
 \caption{90\% C.L. upper limits on the neutrino flux from WIMP annihilations in the Milky Way as a function of the WIMP masses for the different channels considered. For this plot the NFW profile was used.}
\label{Flux}
\end{figure}

The resulting upper limits in terms of neutrino flux are shown in Figure \ref{Flux}. For each annihilation channel and WIMP mass range, the reconstruction strategy, QFit or $\Lambda$Fit, which gives the best sensitivity is used in the final result. $\Lambda$Fit is
used for $\rm M_{WIMP} \geq 260$ $\text{GeV}/\text{c}^2$ for the $\tau^+ \tau^-$ and $\mu^+\mu^-$ channels; for $\rm M_{WIMP} \geq 750$ $\text{GeV}/\text{c}^2$ for the $b \bar b$ channel;
for $\rm M_{WIMP} \geq 150$ $\text{GeV}/\text{c}^2$ for $W^+ W^-$ and for $\rm M_{WIMP} \geq 100$ $\text{GeV}/\text{c}^2$ for the $\nu_\mu \bar{\nu}_\mu$ channel. For the remaining values, i.e at low WIMP masses, the QFit results are used. 

\begin{figure}[h!]
	\centering
	\includegraphics[width=0.49\textwidth]{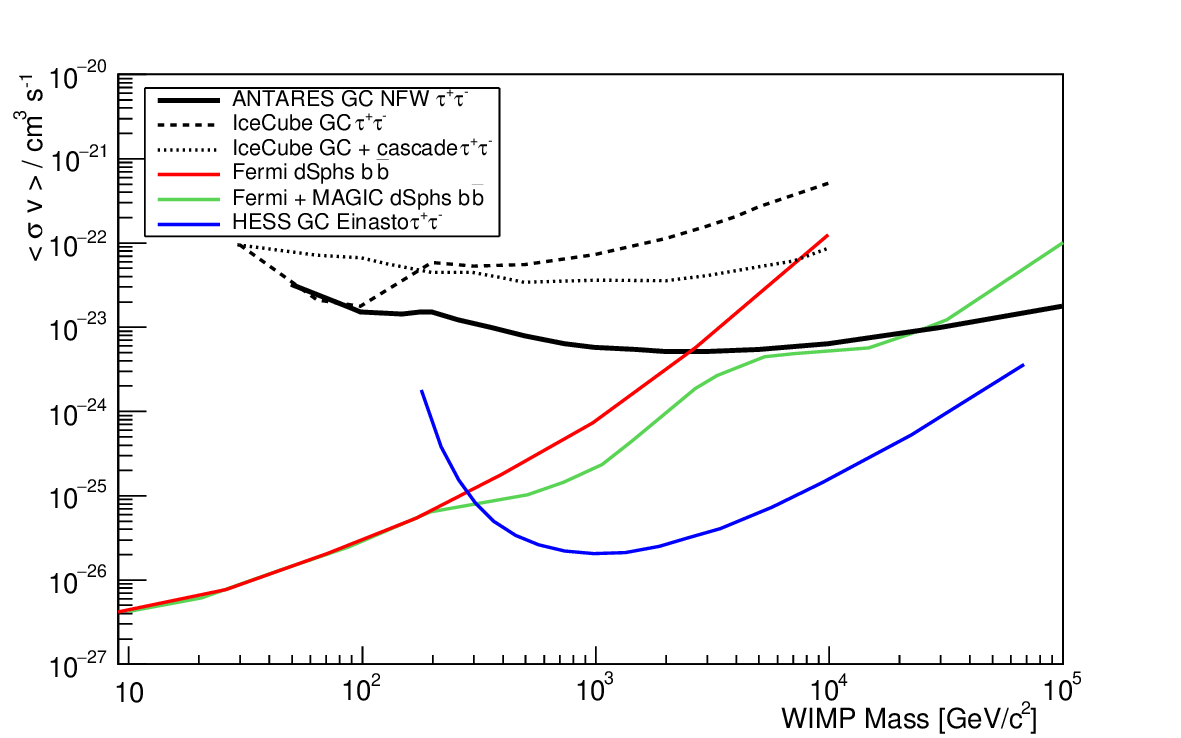} 
	\caption{90\% C.L. limits on the thermally averaged annihilation cross--section, $\langle\sigma \mathrm{v}\rangle$, as a function of the WIMP mass in comparison to the limits from other experiments~\cite{IC_GC,IC_GC_CASC,FERMI_dwarf,FERMI_Mag,HESS_new}. The results from IceCube and ANTARES were obtained with the NFW profile.}
	\label{sv_comp}
\end{figure}

\begin{figure}[h!]
	\centering
	\includegraphics[width=0.49\textwidth]{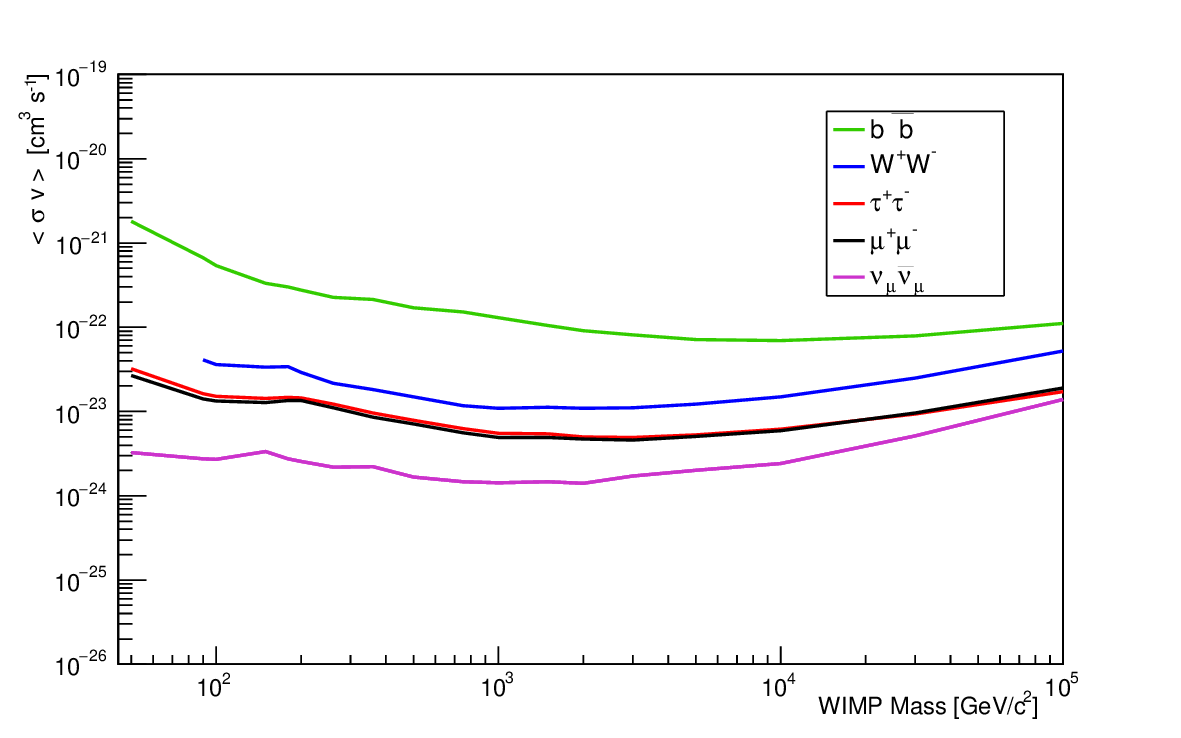} 
	\caption{90\% C.L. limits on the thermally averaged annihilation cross--section, $\langle\sigma \mathrm{v}\rangle$, as a function of the WIMP mass for all annihilation channels using the NFW halo profile.}
	\label{sv_allchannel}
\end{figure}

From the limits on the neutrino flux, limits on $\langle\sigma \mathrm{v}\rangle$ can be derived. The 90\% C.L. upper limit on $\langle\sigma \mathrm{v}\rangle$ for the $\tau^+\tau^-$ channel as a function of the WIMP mass is shown in Figure \ref{sv_comp}, compared with limits obtained by other indirect searches. Most of the direct search experiments are not directly sensitive to $\langle\sigma \mathrm{v}\rangle$. The limits for all annihilation channels for the NFW halo profile are shown in Figure \ref{sv_allchannel}. 

The IceCube results presented in Figure \ref{sv_comp} (using tracks only~\cite{IC_GC} and using cascades as well \cite{IC_GC_CASC}) refer to the same channel and the same halo model, therefore the difference between the limits is due to the detector performance, position and integrated live time. The centre of the Milky Way is above the horizon of the IceCube detector and consequently the neutrino candidates correspond to downgoing events. To select neutrino candidates in the analyses of IceCube a veto for tracks starting outside the central part of the detector has to be used, which reduces the acceptance. This, in addition to the better angular resolution of ANTARES and the larger integrated live time in this analysis, explains the difference between the limits.

For the analysis by H.E.S.S. a different set of halo parameter values is used, leading to a more extended source. The results of FERMI and MAGIC are based on dwarf spheroidal galaxies and use the $\mathrm{b \bar{b}}$ annihilation channel. Results from direct detection experiments are not shown since these experiments are typically not sensitive to $\langle \sigma v \rangle$.

This result allows to partly constrain models where the extraterrestrial neutrinos observed by IceCube are partly explained in terms of annihilating dark matter candidates~\cite{MESE}. For WIMP masses above 100~$\text{GeV}/\text{c}^2$ the limitations from partial-wave unitarity~\cite{unitarity} will become relevant, although there is an approach to overcome these limitations~\cite{profumo}.

\begin{figure}[h!]
\centering
\includegraphics[width=0.49\textwidth]{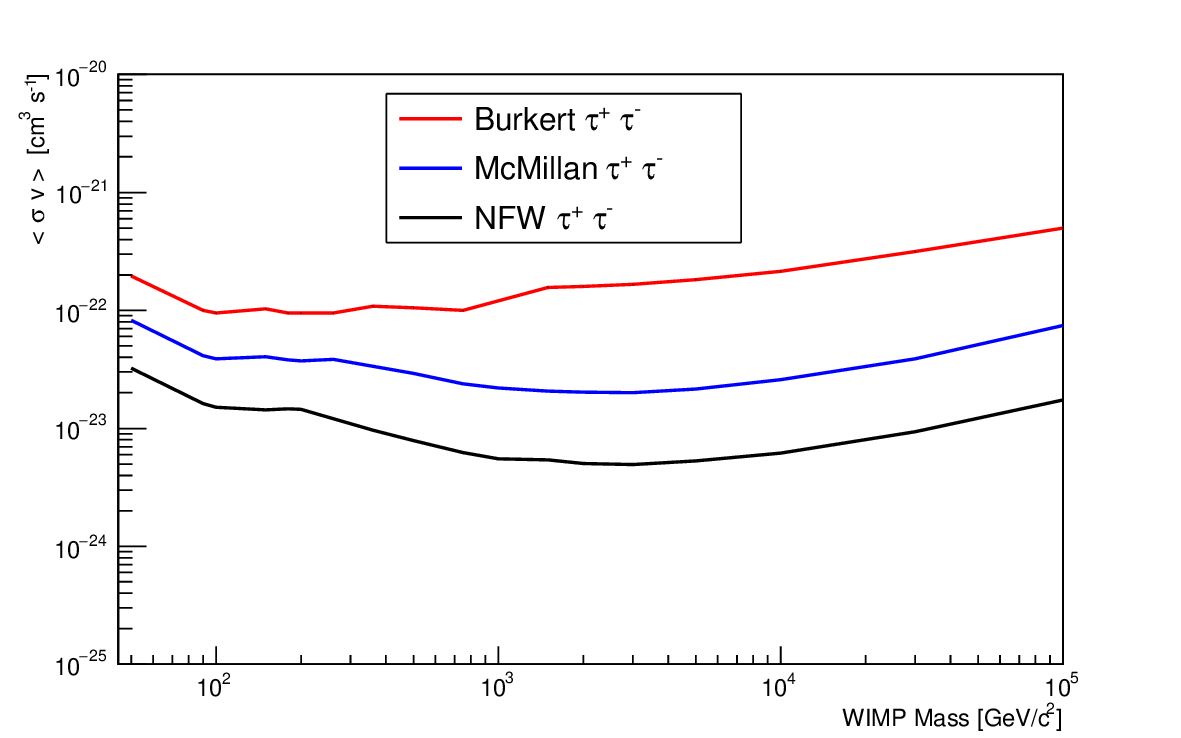} 
 \caption{90\% C.L. limits on the thermally averaged annihilation cross--section, $\langle\sigma \mathrm{v}\rangle$, as a function of the WIMP mass for the three considered halo models for the $\tau^{+}\tau^{-}$ channel.}
\label{sv_model}
\end{figure}

In order to illustrate the large effect of the choice of the halo model and the profile parameters, a comparison between upper limits derived using the NFW, the Burkert and the McMillan results is shown in Figure \ref{sv_model} for the $\tau^{+}\tau^{-}$ channel. As can be seen, depending on the WIMP mass, differences of more than one order of magnitude are observed between the different halo models.

\section{Conclusions}

The results from a new search for dark matter annihilation in the Milky Way using data from the ANTARES neutrino telescope from 2007 to 2015 show no excess above the expected background.
Limits at 90\% C.L. have been set for the NFW, the McMillan and the Burkert profile, five annihilation channels and WIMP masses ranging from 50~$\text{GeV}/\text{c}^2$ to 100~$\text{TeV}/\text{c}^2$. 
In the present version, the 90$\%$ C.L limits reported in Figs. 3 to 6 have been obtained from an integrated acceptance (eq. 8) computed more accurately. 
These upper limits are weaker for WIMP masses larger than 1 TeV/c$^2$ with respect to those computed in the previous version of this paper. 
They remain the most stringent for a certain region of the parameter space arising from neutrino detectors.

\section*{Acknowledgements}

The authors acknowledge the financial support of the funding agencies:
Centre National de la Recherche Scientifique (CNRS), Commissariat \`a
l'\'ener\-gie atomique et aux \'energies alternatives (CEA),
Commission Europ\'eenne (FEDER fund and Marie Curie Program),
Institut Universitaire de France (IUF), IdEx program and UnivEarthS
Labex program at Sorbonne Paris Cit\'e (ANR-10-LABX-0023 and
ANR-11-IDEX-0005-02), Labex OCEVU (ANR-11-LABX-0060) and the
A*MIDEX project (ANR-11-IDEX-0001-02),
R\'egion \^Ile-de-France (DIM-ACAV), R\'egion
Alsace (contrat CPER), R\'egion Provence-Alpes-C\^ote d'Azur,
D\'e\-par\-tement du Var and Ville de La
Seyne-sur-Mer, France;
Bundesministerium f\"ur Bildung und Forschung
(BMBF), Germany; 
Istituto Nazionale di Fisica Nucleare (INFN), Italy;
Stichting voor Fundamenteel Onderzoek der Materie (FOM), Nederlandse
organisatie voor Wetenschappelijk Onderzoek (NWO), the Netherlands;
Council of the President of the Russian Federation for young
scientists and leading scientific schools supporting grants, Russia;
National Authority for Scientific Research (ANCS), Romania;
Mi\-nis\-te\-rio de Econom\'{\i}a y Competitividad (MINECO):
Plan Estatal de Investigaci\'{o}n (refs. FPA2015-65150-C3-1-P, -2-P and -3-P, (MINECO/FEDER)), Severo Ochoa Centre of Excellence and MultiDark Consolider (MINECO), and Prometeo and Grisol\'{i}a programs (Generalitat
Valenciana), Spain; 
Ministry of Higher Education, Scientific Research and Professional Training, Morocco.
We also acknowledge the technical support of Ifremer, AIM and Foselev Marine
for the sea operation and the CC-IN2P3 for the computing facilities.

\appendix


\bibliography{manuscript}

\end{document}